\documentclass[10pt,letterpaper]{article}
\usepackage[top=0.85in,left=1.25in,footskip=0.75in]{geometry}

\usepackage{amsmath,amssymb}

\usepackage{changepage}

\usepackage[utf8x]{inputenc}

\usepackage{textcomp,marvosym}

\usepackage{cite}

\usepackage{nameref,hyperref}

\usepackage[right]{lineno}
\usepackage{ragged2e}

\usepackage{microtype}
\DisableLigatures[f]{encoding = *, family = * }

\usepackage[table]{xcolor}

\usepackage{array}

\newcolumntype{+}{!{\vrule width 2pt}}

\newlength\savedwidth

\usepackage{booktabs}       
\usepackage{amssymb}
\usepackage{amsmath}
\usepackage{amsfonts}   
\usepackage{subcaption} 

\usepackage{algorithm}
\usepackage{algpseudocode}

\usepackage{setspace} 
\raggedright
\usepackage[aboveskip=1pt,labelfont=bf,labelsep=period,singlelinecheck=off]{caption}

\makeatletter
\renewcommand{\@biblabel}[1]{\quad#1.}
\makeatother

\usepackage{lastpage,fancyhdr,graphicx}
\usepackage{epstopdf}
\fancyheadoffset[L]{2.25in}
\usepackage{tabularx}

\newcommand{\markchanges}[1]{#1}

\newcommand{\markchangesnew}[1]{#1}

\newcommand{\given}{\,|\,}
\graphicspath{{figures/}} 
\DeclareMathOperator*{\argmin}{argmin}

\usepackage{ifthen}

\newcommand{\hidefigures}{false}
\ifthenelse{\equal{\hidefigures}{true}}
{
    \renewcommand{\includegraphics}[2][]{}
}

\begin{document}
\vspace*{0.2in}

% Title must be 250 characters or less.
\begin{flushleft}
{\Large
\textbf\newline{OutbreakFlow: Model-based Bayesian inference of disease outbreak dynamics with invertible neural networks
and its application to the COVID-19 pandemics in Germany} % Please use "sentence case" for title and headings (capitalize only the first word in a title (or heading), the first word in a subtitle (or subheading), and any proper nouns).
}
\newline
% Insert author names, affiliations and corresponding author email (do not include titles, positions, or degrees).
\\
Stefan T. Radev\textsuperscript{1*},
Frederik Graw\textsuperscript{2},
Simiao Chen\textsuperscript{3,4},
Nico T. Mutters\textsuperscript{5},
Vanessa M. Eichel\textsuperscript{6},
Till B\"{a}rnighausen\textsuperscript{3,7,8},
Ullrich K\"{o}the\textsuperscript{9},
\\
\bigskip
\textbf{1} Institute of Psychology, Heidelberg University, Heidelberg, Germany\\
\textbf{2} BioQuant - Center for Quantitative Biology, Heidelberg University, Heidelberg, Germany \\
\textbf{3} Heidelberg Institute of Global Health, Heidelberg, Germany \\
\textbf{4} Chinese Academy of Medical Sciences and Peking Union Medical College, Beijing, China\\
\textbf{5} Institute for Hygiene and Public Health, University Hospital Bonn, Bonn, Germany\\
\textbf{6} Center of Infectious Diseases, University Hospital Heidelberg,  Heidelberg, Germany\\
\textbf{7} Harvard T.H. Chan School of Public Health, Boston, Massachusetts, USA\\
\textbf{8} Africa Health Research Institute, Durban, South Africa\\
\textbf{9} Computer Vision and Learning Lab, Heidelberg University, Heidelberg, Germany\\

\bigskip

% Insert additional author notes using the symbols described below. Insert symbol callouts after author names as necessary.
% 
% Remove or comment out the author notes below if they aren't used.
%
% Primary Equal Contribution Note
%\Yinyang These authors contributed equally to this work.

% Additional Equal Contribution Note
% Also use this double-dagger symbol for special authorship notes, such as senior authorship.
%\ddag These authors also contributed equally to this work.

% Current address notes
%\textcurrency Current Address: Dept/Program/Center, Institution Name, City, State, Country % change symbol to "\textcurrency a" if more than one current address note
% \textcurrency b Insert second current address 
% \textcurrency c Insert third current address

% Deceased author note
%\dag Deceased

% Group/Consortium Author Note
%\textpilcrow Membership list can be found in the Acknowledgments section.

% Use the asterisk to denote corresponding authorship and provide email address in note below.
* stefan.radev93@gmail.com

\end{flushleft}
% Please keep the abstract below 300 words

\justifying

\section*{Abstract}
Mathematical models in epidemiology are an indispensable tool to determine the dynamics and important characteristics of infectious diseases. Apart from their scientific merit, these models are often used to inform political decisions and interventional measures during an ongoing outbreak. 
However, reliably inferring the epidemical dynamics by connecting complex models to real data is still hard and requires either laborious manual parameter fitting or expensive optimization methods which have to be repeated from scratch for every application of a given model.
In this work, we address this problem with a novel combination of epidemiological modeling with specialized neural networks.
Our approach entails two computational phases:
In an initial training phase, a mathematical model describing the epidemic is used as a coach for a neural network, which acquires global knowledge about the full range of possible disease dynamics. 
In the subsequent inference phase, the trained neural network processes the observed data of an actual outbreak and infers the parameters of the model in order to realistically reproduce the observed dynamics and reliably predict future progression. With its flexible framework, our \markchanges{simulation-based} approach is applicable to a variety of epidemiological models.
Moreover, since our method is fully Bayesian, it is designed to incorporate all available prior knowledge about plausible parameter values and returns complete joint posterior distributions over these parameters.
Application of our method to the early Covid-19 outbreak phase in Germany demonstrates that we are able to obtain reliable probabilistic estimates for important disease characteristics, such as generation time, fraction of undetected infections, likelihood of transmission before symptom onset, and reporting delays using a very moderate amount of real-world observations. 

% Please keep the Author Summary between 150 and 200 words
% Use first person. PLOS ONE authors please skip this step. 
% Author Summary not valid for PLOS ONE submissions.   
\section*{Author summary}
Emerging infections and epidemic outbreaks are associated with large uncertainties concerning data integrity that challenge the timely detection of disease characteristics and dynamics. 
Robust parameter inference for mathematical models aiming to describe these dynamics is essential to predict the progression of an epidemic and inform on appropriate public health interventions. 
In this study, we present a novel method based on invertible neural networks that allows inference of important epidemiological characteristics in case of limited data, thereby allowing for reliable uncertainty quantification. 
The method circumvents common challenges associated with sparse data by using simulation-based training of an expressive generative neural network.
Applying our method to data of the early Covid-19 epidemic in Germany, we are able to obtain reliable estimates on important disease characteristics, such as the proportion of \markchanges{infected individuals remaining undetected}, despite limited observations during early outbreak dynamics.

\nolinenumbers

% Use "Eq" instead of "Equation" for equation citations.
\section*{Introduction}
Assessing important disease characteristics and transmission dynamics is of utmost importance in the case of new epidemic outbreaks in order to forecast their progression and guide effective public health measurements. 
Mathematical models that provide a reliable representation of the processes driving the dynamics of an epidemic are an essential tool for this task (see for example \cite{keeling2011modeling}). In the case of communicable diseases, these models typically take the form of systems of ordinary differential equations governing the transitions between different population compartments, such as, ``Susceptible'', ``Infected'', and ``Recovered'' (SIR)  \cite{keeling2011modeling}. 
Provided that intrinsic properties of the disease (e.g., transmission rates and recovery periods) are already known, SIR models and their extensions are successfully used to simulate outcomes of possible public health interventions.

% Covid 19 - importance of parameter inference
However, for newly emerging infectious diseases, such as Covid-19, most of these properties are initially unknown and must be estimated before realistic predictions can be made. 
\markchanges{The task of determining these properties is additionally hampered by limited data availability and integrity within the early outbreak phase.}
During the initial outbreak of the Covid-19 pandemic, model-based inference was used to provide rapid estimates of key epidemiological parameters, which otherwise can be difficult to infer directly from primary clinical tracing data. 
For instance, an earlier study \cite{wu2020nowcasting} incorporated domestic and international travel from and to Wuhan city in a SEIR model and used reported cases outside of Wuhan to infer the reproduction number $R_0$ and epidemic doubling time. 
Similar approaches were used to estimate the reproduction number of Covid-19 in various other settings \cite{chinazzi2020effect,kucharski2020early,scire2020reproductive}. 
Other studies aimed at identifying critical epidemiological characteristics, such as age-specific mortality rates \cite{hauser2020estimation}, the impact of implemented control measures on disease transmission \cite{tian2020investigation,dehning2020inferring}, or the fraction of undocumented infections \cite{li2020substantial}.
%TODO - add here some common problem of those studies, problem is, that we also cite dehning and tian, who also used complex models and Bayesian methods. I guess we should pitch the deep learning approach...

% Problem and approach
Importantly, since such SIR-type models are employed to forecast the dynamics of an epidemic dependent on public interventions or seasonal effects, reliable inference of such key epidemiological parameters and trustworthy uncertainty quantification is paramount to support decision making. 
The estimation of hidden model parameters from observations of model outcomes (\textit{inverse inference}) is also referred to as \textit{model calibration} in the medical decision and health policy modeling literature \cite{kong2009calibration}. 
% The goal of inverse inference is to determine the values of the unknown model parameters such that the model outputs match the observed real-world data, the so-called \textit{calibration targets}, as close as possible. 
Traditionally, model calibration has been considered as an optimization problem seeking the best possible parameter configuration \markchanges{explaining the data} (e.g., by performing non-linear least squares minimization \markchanges{or relying on maximum likelihood estimation \cite{martcheva2015introduction}}). 
However, \markchangesnew{the resulting optimization and maximum likelihood methods focus on point estimates for the individual parameters and usually lack appropriate approaches to assess their accuracy.
This is a severe disadvantage, because reliable uncertainty quantification is crucial when the estimates shall be used to model and predict} future outcomes.

% Bayesian approach
In contrast to optimization and \markchanges{maximum likelihood} approaches, Bayesian methods provide a principled way to quantify the \markchanges{uncertainty of inferred model parameters}, as they return full posterior distributions for the unknown parameters rather than single point estimates. 
Markov chain Monte Carlo (MCMC) sampling represents \markchanges{one of the classical approaches to Bayesian model calibration \cite{gelman2013bayesian}, and it has been extensively used in Covid-19 studies to infer model parameters describing the dynamics of the disease \cite{chinazzi2020effect, peak2020individual, kucharski2020early, dehning2020inferring}.}
However, Bayesian model calibration is computationally expensive and depends explicitly on the evaluation of the likelihood function of model parameters.
When the likelihood function is intractable or unknown, approximate Bayesian computation (ABC) can be used to approximate the posterior distribution of parameters \cite{minter2019approximate, menzies2017bayesian}.
However, standard ABC methods notoriously suffer from poor scalability \markchangesnew{(i.e., the efficient application to large data sets and complex models)}, which confines their utility to relatively simple models containing only a few parameters \cite{cranmer2020frontier}.

% Our approach
\markchanges{In order to overcome the limitations of individual parameter estimation methods,} our approach aims to combine the advantages of optimization-based and Bayesian methods by using specialized invertible neural networks.
In particular, we develop a novel methodological framework based on a neural network architecture \markchanges{called BayesFlow \cite{radev2020bayesflow} to facilitate model-based inference with a primary, but not exclusive, focus on epidemiology.}
Our method can incorporate an arbitrary number of epidemiological time series (or other type of temporal information) \markchanges{and can, in principle, be applied to any dynamic model described by (stochastic) ordinary differential equations.
Moreover, since the length of the time series is not pre-determined during training and inference, we can \markchangesnew{consider additional information in our analyzes without having to re-train the networks.}
By using extensive simulation-based training, our method circumvents the general necessity for large training sets that are lacking during emerging epidemics. 
Additionally, our method returns posterior distributions which are fully compatible with a Bayesian interpretation and can thus be used to assess the uncertainty associated with any estimation and prediction quantity.}

% Results & Outcome
We demonstrate the feasibility of our method by analyzing public Covid-19 data for Germany and the individual German federal states based on the \markchanges{reported daily number of infected, recovered and deceased cases during the first months of the pandemic. }
Our neural network is trained using simulations from a customized SEIR-model variant \cite{khailaie2020estimate}, in combination with an observation model accounting for the differences between true and reported case numbers, and an intervention model describing the \markchanges{intervention measures for prevention and control} imposed by German authorities \cite{dehning2020inferring}. 
Despite the limited number of measurements and the considerable complexity of the model with 34 unknown parameters in total, \markchangesnew{our network manages to extract information for more than half of the parameters}. Credibility intervals of our parameter estimates are well in line with independently published results, and \markchanges{re-simulations starting at our estimated parameters reproduce the observed time series very well}. In particular, our inference suggests that approximately four fifths of all infectious individuals remained undetected across all German federal states.

\section*{Materials and Methods}
\subsection*{Data}

The model was applied to epidemiological data on the number of reported Covid-19 cases (infected, recovered and deceased) in Germany and the individual federal states \markchanges{(only infected and diseased)} from March 01, 2020 until June 11, 2020. 
Data were collected from publicly available sources \markchanges{during the same time period and were not subsequently cleaned or corrected in the aftermath.
Therefore, all sources of uncertainty remain in the data as they would have in the early days of an ongoing pandemic.} Code and scripts for reproducing all results and figures as well as \markchanges{the general framework of OutbreakFlow} for training new networks on new models are available at \url{https://github.com/stefanradev93/AIAgainstCorona}.

\subsection*{Neural Bayesian Parameter Estimation}

Following a Bayesian approach for parameter estimation requires prior knowledge about reasonable parameter ranges \cite{gelman2013bayesian}. 
Combining this prior knowledge with information extracted from the observed data leads to a posterior distribution which is generally narrower than the prior and thus expresses our updated state of knowledge and associated uncertainty for the individual parameters.
More formally, let $\theta$ be the vector of all unknown model parameters and $X := x_{1:T} = [x_1,...,x_T]$ a multivariate epidemiological time series with $T$ individual time points indicating, for instance, the number of infected, recovered and diseased individuals.
Then the well-known analytical formula for the posterior according to Bayes' rule is
\begin{equation}
    p(\theta \given X) = \frac{p(X \given \theta)\,p(\theta)}{\int p(X \given \theta)\,p(\theta)\,d \theta} \label{eq:2}
\end{equation}
where $p(X \given \theta)$ represents the likelihood of observing data $X$ when the true parameters are $\theta$, $p(\theta)$ is the prior distribution encoding our knowledge about plausible parameter configurations for $\theta$, and the denominator represents a normalizing constant (usually called the evidence).

Despite being conceptually simple, this formula poses two major challenges in the present setting:
(i) Efficient and accurate approximation of the intractable posterior $p(\theta \given X)$ is challenging;
(ii) The likelihood is only implicitly defined via realizations $X \sim p(X \given \theta)$ generated by repeatedly running \markchanges{simulations of the underlying epidemiological model with different $\theta$.}

We solve both problems with our recently proposed neural Bayesian inference architecture \markchanges{called BayesFlow, which is explained in full mathematical details in the corresponding methodological work \cite{radev2020bayesflow}.
The core component of BayesFlow is an {\it invertible neural network} which enables a bidirectional flow of information.}
During the training phase, the invertible network is run in forward direction to learn an accurate model $q(\theta \given X)\approx p(\theta \given X)$ for the posterior distribution of parameters $\theta$ given observations $X$, using a large number of simulated pairs $(X_i,\theta_i)\sim p(X\given\theta)\,p(\theta)$ as training data. 
During the inference phase, the network \markchanges{makes use of its invertible architecture and operates in the inverse direction to estimate the posterior $q(\theta\given X=x^{\text{obs}})\approx p(\theta\given X=x^{\text{obs}})$ for the {\it actually} observed data $x^{\text{obs}}$.}

\markchanges{Validation experiments reported in \cite{radev2020bayesflow} have demonstrated for various model systems, that the BayesFlow method (i) can estimate complex stochastic models of widely varying data types and sizes (e.g., population time series, predator-pray population time series, human response-time data); (ii) outperforms variational or dropout methods for uncertainty quantification; (iii) learns data representations which are more informative than manually selected summary statistics; and (iv) outperforms case-based methods such as ABC and MCMC, whose computations must be re-run from scratch for every observed dataset.

The latter advantage is called ``amortized inference'': a BayesFlow network learns to generalize the training knowledge and can efficiently apply it to multiple real observations without retraining.
The network's training effort thus quickly \textit{amortizes} over a sequence of inference queries (e.g., time series), in contrast to sampling methods (e.g., MCMC), which cannot leverage experience and require the same large computational effort for every query.
In addition, fast amortized inference facilitates model validation by enabling efficient probabilistic calibration and posterior predictive checks on large validation datasets \cite{radev2020bayesflow}.}

\subsection*{OutbreakFlow -- The BayesFlow Approach to Epidemiological Inference}

\markchanges{We propose OutbreakFlow, an instantiation of our BayesFlow architecture that utilizes a novel combination of three jointly trained neural modules to analyze noisy multi-variate time series with potentially long-term temporal dependencies, as are typical in the context of epidemiology.
It can process both short and long time series and can thus perform efficient Bayesian updating as new data become available (e.g., on a daily basis in case of Covid-19).  
Moreover, our method can incorporate additional prior knowledge in the formulation of the underlying generative model.}

Our neural architecture comprises three sub-networks: (i) a convolutional \textit{filtering} network performing noise reduction and feature extraction on the raw observational data; (ii) a recurrent \textit{summary} network reducing \markchanges{filtered time series of {\it arbitrary}} length to statistical summaries of {\it fixed} size; and (iii) an invertible \textit{inference} network performing Bayesian parameter inference, given the learned summaries of the observations.
\markchanges{Fig 1 depicts the training and inference phase with our inference architecture and its essential elements.}

\begin{figure*}
\centering
\includegraphics[width=1.0\textwidth]{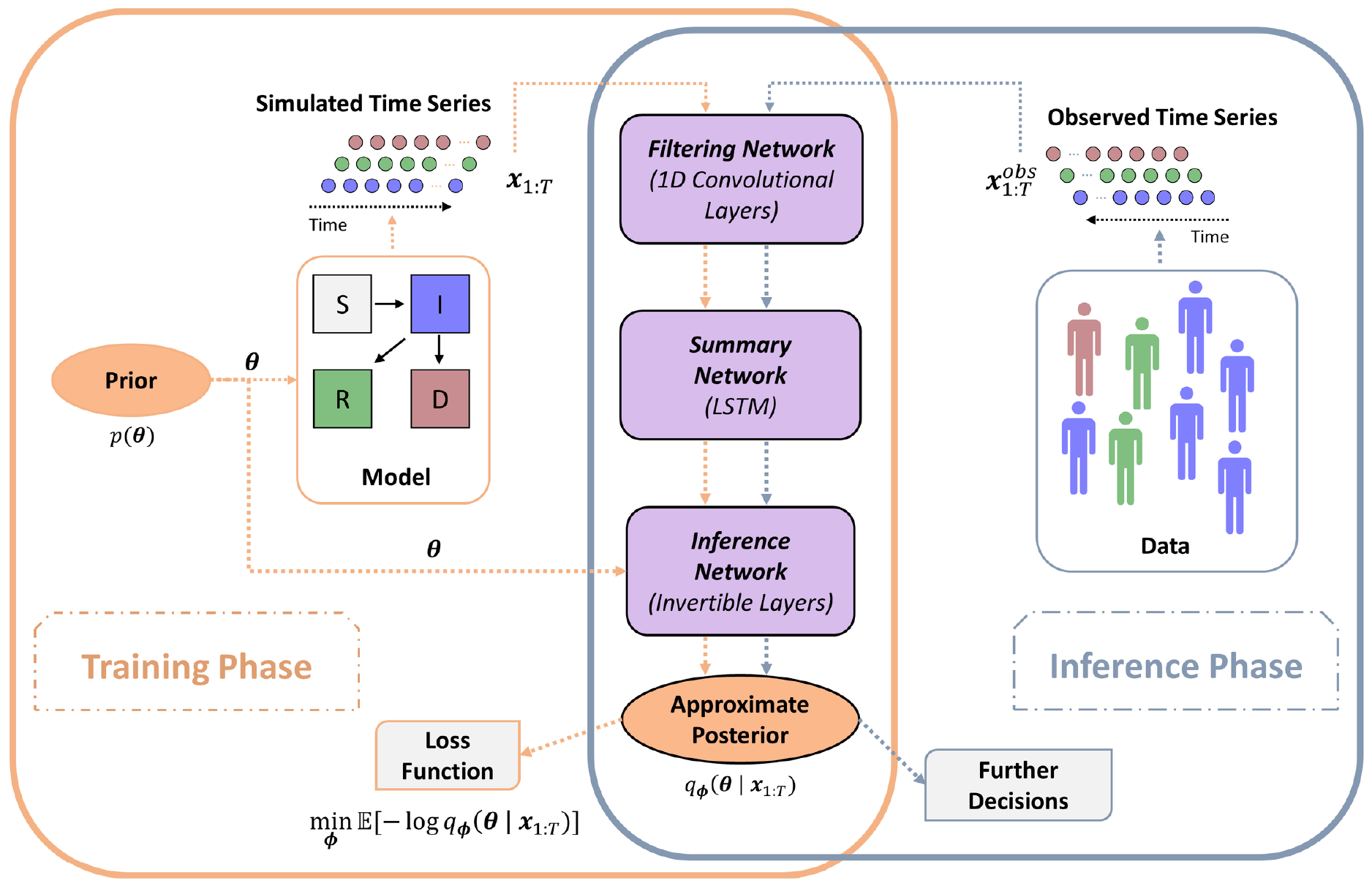}
\caption{\markchanges{Structure and workflow of OutbreakFlow. During the training phase (orange frame on the left), the assumed epidemiological model is used to simulate time series resembling the observed epidemiological data, based on prior distributions of the unknown parameters. The synthetic time series are used to train the composite neural network consisting of (i) a convolutional filtering network, which extracts relevant features while preserving the temporal structure of the data, (ii) a summary network, which reduces the transformed time series to a fixed-sized vector of maximally informative representations, and (iii) an inference network, which estimates the joint parameter posterior from these data representations. During the inference phase (blue frame on the right), the real epidemiological data $\mathbf{x}^{obs}_{1:T}$ are passed to the trained network to infer the posterior distribution of the unknown disease parameters. A full description of the architecture and the methodology is provided within {\bf Materials and Methods}.}}
\label{fig:bayesflow}  
\end{figure*}

The design of the convolutional network is inspired by the well-known Inception network, which has shown tremendous success in computer vision tasks \cite{szegedy2015going}. 
In particular, our network is implemented as a deep fully convolutional network which applies adjustable one-dimensional filters of different size at each level (cf. Fig 1). 
The intuition behind this design is that filters of different size might capture patterns at different temporal scales (e.g., a filter of size one will capture daily fluctuations whereas a filter of size seven will capture weekly dynamics). 
This, in turn, should ease the task of extracting informative temporal features for parameter estimation.

The output of the convolutional network is a multivariate sequence containing the filtered epidemiological time series. 
In order to reduce the filtered sequence to a fixed-size vector, we pass it through a long-short term memory (LSTM) recurrent network \cite{gers2000learning}.
\markchanges{In contrast to standard feed-forward neural networks, LSTM networks incorporate feedback connections which make them ideally suited for processing sequences of observations such as time series.}

\markchanges{A standard LSTM network consists of a cell and three gates.
The cell stores the \textit{internal memory} of the network over arbitrary temporal intervals.
The three gates, comprising an input gate, an output gate, and a forget gate, interact in determining the importance of old and new information.
Importantly, LSTM networks can deal with sequences of varying length, which enables them to process data whose duration is dictated by data availability and to perform online inference, i.e. process new data instantly as they become available.
In contrast to predefined pooling operations (e.g., mean, max, or variance), our recurrent networks learn pooling operations that are adapted to the data and can thus extract potentially much richer representations.}
In this way, our inference architecture learns to filter and extract the most informative features from the noisy observations in an end-to-end manner.
\markchanges{Thus, the user is freed from the difficult (and usually suboptimal) task to hand-engineer suitable data features (summary statistics).}
Finally, the inference network has the task of \textit{inverting} the forward model given the information extracted by the convolutional and recurrent networks (see also \cite{radev2020bayesflow} for more details on the design of invertible networks for inference). 

The invertible network has two modes of operation. During training, the network is only evaluated in the \textit{forward direction} and encouraged via a suitable optimization criterion to transform the posterior into a simple base distribution (e.g., Gaussian) from which samples can be easily obtained.
Thus, the inference network integrates information from both the prior and the data-generating mechanism (i.e., the implicit likelihood).

During inference, the inference network is only evaluated in the \textit{inverse direction} using conditional information from real observed data passed through the filtering and summary networks. 
The posterior is approximated by repeatedly sampling from the simple base distribution and applying the inverse of the forward transformation learned during the training phase. Importantly, this method recovers the true posterior under perfect convergence of the optimization method \cite{radev2020bayesflow}.

More formally, let us denote the functions represented by the three networks as $g$, $h$, and $f$. 
Then the filtering network determines a filtered time series $\widetilde{x}_{1:T}=g(x_{1:T})$ from observed data $x_{1:T}$, where the number of time steps $T$ depends on data availability.
The summary network turns the output of the filtering network into a fixed-size representation $y=h(\widetilde{x}_{1:T})$ by keeping only the final output vector of the LSTM network, which encodes the accumulated information over all observed time steps.
Finally, the inference network generates samples $\widehat{\theta}\sim q(\theta\given x_{1:T})$ from the parameter posterior by computing $\widehat{\theta}=f^{-1}(y, z)$ with normally distributed random vectors $z\sim \mathcal{N}(0,\mathbb{I})$.  
The parameters of all three networks are optimized jointly during the training phase. Denoting the vector of all trainable network parameters as $\phi$, the three networks solve the following optimization criterion
\begin{linenomath*}
\begin{align}
    \widehat{\phi} &= \argmin_{ \phi}\,\,\mathbb{E}_{X \sim p(X)}\big[\mathbb{KL}(p(\theta\given X)\,||\,q_{\phi}(\theta\given X))\big] \\
    &= \argmin_{\phi} \,\,\mathbb{E}_{(X,\theta) \sim p(X,\theta)}\big[-\log q_{\phi}(\theta\given X)\big] 
    \label{eq:nll}
\end{align}
\end{linenomath*}
where $\mathbb{KL}(p \,||\, q)$ denotes the Kullback-Leibler divergence \cite{van2014renyi} between probability density functions $p$ and $q$.
We approximate the latter expectation via its empirical mean over realizations $(X,\theta) \sim p(\theta, X)$ obtained via simulations from a forward model.

As previously mentioned, one of the most important advantages of our method is \textit{amortized inference}, owing to the fact that we approximate the posterior \textit{globally} via a single set of network parameters $\widehat{\phi}$. 
This is especially advantageous in epidemiological contexts, where the same model is applied in multiple populations (countries, cultures) or at different scales (states, regions), since the same pre-trained model can be repeatedly utilized for different populations and scales. 
Indeed, in the following real-world application, we demonstrate efficient amortized inference and excellent predictive performance with a single architecture applied simultaneously to epidemiological data from all German federal states.  

\subsection*{The Epidemiological Model}
\label{sec:epidemiological-model}

In order to account for the specific nature of the current Covid-19 outbreak, our epidemiological model consists of three submodels: 
(i) a disease model describing the true dynamics of relevant population compartments;
(ii) an intervention model describing the strengthening and relaxation of non-pharmaceutical intervention measures; and
(iii) an observation model describing the deviations of publicly reported data from the true values.
These models build upon the previous work of Khailaie et al. \cite{khailaie2020estimate} and Dehning et al. \cite{dehning2020inferring}, who adapted the general SIR approach to the specifics of the Covid-19 pandemic and the situation in Germany.
Parameter priors are based on our current state of knowledge about disease characteristics and government measures, but are chosen very wide to prevent them from dominating the information extracted from the actual observations.

\noindent
\\{\bf Disease Model:} The disease model is a system of non-linear ordinary differential equations (ODEs) distinguishing between six compartments: 
susceptible ($S$), exposed ($E$ - infected individuals who do not show symptoms and are not yet infectious), infected ($I$ - symptomatic cases that are infectious), carrier ($C$ - infectious individuals who recover without being detected), recovered ($R$), and dead ($D$), see Fig 2. 
Note that direct recovery from the carrier state $C$ covers all reasons why an infection might go undetected, including, among others, truly asymptomatic cases, lack of follow-up on pre-symptomatic cases, limited testing capacity under minor symptoms -- our model does not differentiate between these posibilities.
Observations with limited accuracy (as described by the observation model below) are available for the compartments $I$, $R$, and $D$.
The true time series of all compartments are therefore considered latent and need to be estimated.

The ODEs for our epidemiological model are defined by:
\begin{linenomath*}
\begin{align}
    \frac{dS}{dt} &= -\lambda (t)\,\left(\frac{C + \beta\,I}{N}\right)\,S \\
    \frac{dE}{dt} &= \lambda (t)\,\left(\frac{C + \beta\,I}{N}\right)\,S - \gamma\,E \\
    \frac{dC}{dt} &= \gamma\,E - (1 - \alpha)\,\eta\,C - \alpha\,\theta\,C\\
    \frac{dI}{dt} &= (1 - \alpha)\,\eta\,C - (1-\delta)\,\mu\,I - \delta\,d\,I \\
    \frac{dR}{dt} &= \alpha\,\theta\,C + (1-\delta)\,\mu\,I \\
    \frac{dD}{dt} &= \delta\,d\,I
\end{align}
\end{linenomath*}
The meaning of the model parameters and their priors are detailed in Table S1. 
Prior ranges are based on considerations in \cite{dehning2020inferring} and \cite{tang2020estimation}. 
All rate parameters are considered to be constant, except for the transmission rate $\lambda(t)$ which is considered to be time-dependent as it accounts for possible behavioral changes implied by non-pharmaceutical interventions.

\noindent
\\{\bf Intervention Model:} The intervention model accounts for changes in the transmission rate $\lambda(t)$ due to non-pharmaceutical interventions and mitigation measures.
Corresponding to the approach followed by \cite{dehning2020inferring}, we define three change points for $\lambda(t)$ encoding an assumed transmission rate reduction in response to intervention measures imposed by the German authorities.
Each change point is represented by a piece-wise linear function with three degrees of freedom: the effect strength and the boundaries defining the time interval for the effect to fully manifest itself.
The chosen priors express the expected effect of each measure to reduce the transmission rate roughly by half after the date when it comes into force, \markchanges{but we allow for wide uncertainty margins to facilitate learning of the actual behavior.}
In addition to the previous approach in \cite{dehning2020inferring}, our model also includes a fourth change point expressing the assumption that an eventual withdrawal of effective intervention measures (officially or due to non-compliance) will lead to a slight increase of the transmission rate.
Note that we assume that interventions do not affect the risk of infection upon contact with a detected infectious individual ($\beta$). 
Prior distributions and descriptions for all parameters are given in Table S2.

\begin{figure}
\centering
\includegraphics[width=\textwidth]{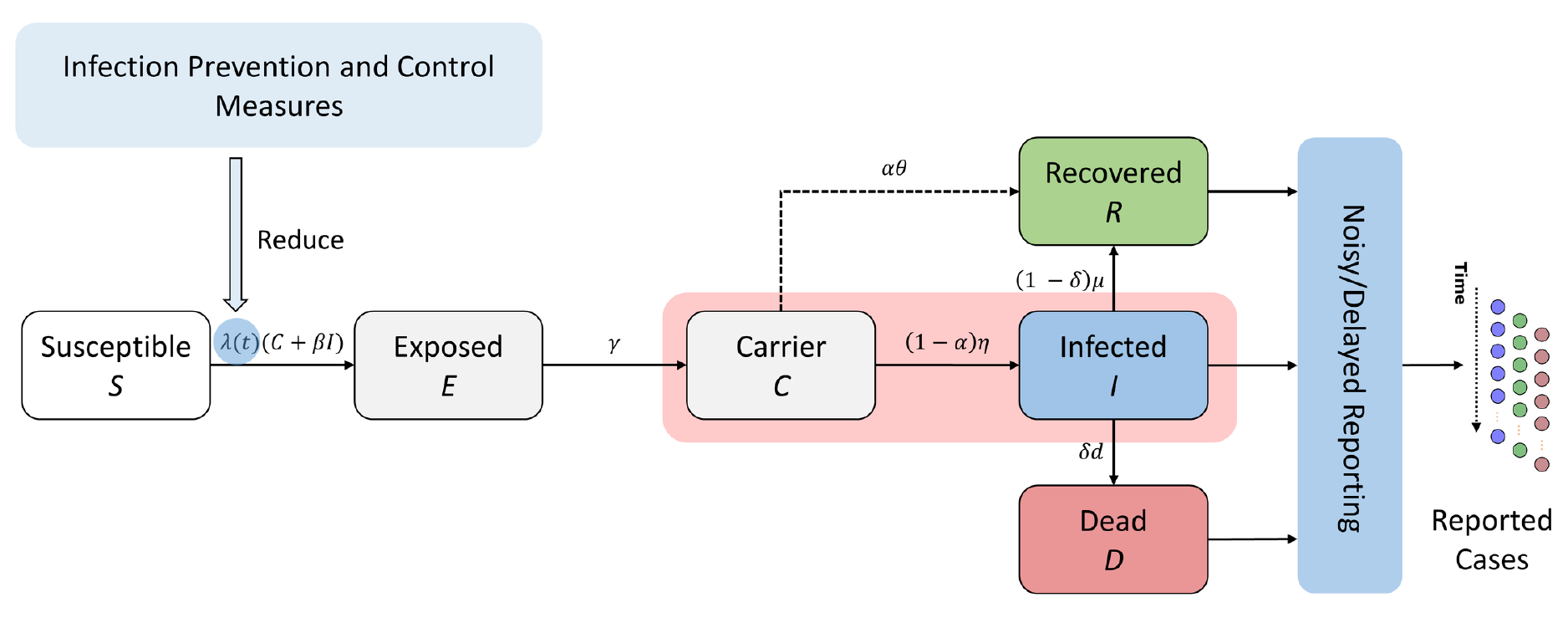}
\caption{\markchanges{Sketch of the mathematical model describing the epidemiological dynamics of Covid-19. The model is of SEIR-type with six compartments: susceptible ($S$), exposed ($E$, i.e. infected but non-infectious), carrier ($C$, i.e. infectious but undetected), infected ($I$, i.e. infectious and diagnosed), recovered ($R$) and dead ($D$) individuals. In addition, the blue boxes indicate model extensions accounting for external factors, namely intervention measures affecting the transmission rate $\lambda(t)$ and imperfect case reporting due to noise or delays. Note, that data is only reported for the observable compartments $I$, $R$, and $D$. For a detailed description of the model and the different components see {\bf Materials and Methods}.}}
\label{fig:disease-model}
\end{figure}

\noindent
\\{\bf Observation Model:}
The observation model accounts for the fact that officially reported cases might not represent the true case numbers of the epidemics. 
It represents three error sources: a delay between actual infection and reporting, the weekly modulation of reporting rates (since testing and reporting activities are considerably reduced on weekends), and a noise term describing random fluctuations.
Separate parameter sets are learned for each of the three publicly reported time series $I^{(obs)}$, $R^{(obs)}$, and $D^{(obs)}$ -- the remaining compartments are considered unobservable. 
The relationship between the reported counts and their true values is described by the following set of discrete-time difference equations with time steps $t$ measured in days.
\begin{linenomath*}
\begin{align}
     I^{(obs)}_t &= I^{(obs)}_{t-1} + (1 - f_I(t))\,(1 - \alpha)\,\eta\,C_{t-L_I} + \sqrt{I^{(obs)}_{t-1}} \,\sigma_I\,\xi_t \\ 
    R^{(obs)}_t &= R^{(obs)}_{t-1} + (1 - f_R(t))\,(1-\delta)\,\mu\,I_{t-L_R}  + \sqrt{R^{(obs)}_{t-1}}\,\sigma_R\,\xi_t \\
    D^{(obs)}_t &= D^{(obs)}_{t-1} + (1 - f_D(t))\, \delta\,d\,I_{t-L_D} + \sqrt{D^{(obs)}_{t-1}}\,\sigma_D\, \xi_t 
\end{align}
\end{linenomath*}
\markchanges{where $L_I, L_R$, and $L_D$ denote the reporting delays (lags), and denote $\sigma_I, \sigma_R$ and $\sigma_D$ the scales of multiplicative reporting noise for the respective compartments.} 
The noise variables $\xi_t$ follow a Student-\textit{t} distribution with 4 degrees of freedom. 
The weekly modulation of reporting coverage $f_{\mathcal{C}}(t)$ for each of the compartments $\mathcal{C} \in \{I, R, D\}$ is computed as follows:
\begin{linenomath*}
\begin{align}
    f_{\mathcal{C}}(t) = (1 - A_{\mathcal{C}})\,\left(1 - \left| \sin \left( \frac{\pi}{7}t - 0.5\, \Phi_{\mathcal{C}} \right) \right|  \right)
\end{align}
\end{linenomath*}
This yields three additional unknown parameters for the weekly modulation amplitudes $A_I, A_R, A_D$, and phases $\Phi_I, \Phi_R, \Phi_D$, each. 
The prior distributions and descriptions for the observation model's parameters are listed in Table S3.

\subsection*{\markchanges{Neural Network Training}}

\markchanges{An OutbreakFlow is trained with simulated data by minimizing the negative log posterior according to Eq.\ref{eq:nll}.
The training phase can be realized in different ways, depending on the modeling scenario and the modelers’ computational budget.
First of all, when only a single time series has to be analyzed, non-amortized methods like \cite{papamakarios2019sequential,greenberg2019automatic} may outperform OutbreakFlow, because they constrain the simulation scope to the vicinity of the observed data.
On the other hand, when the model has to be applied to {\em multiple} observed time series (e.g., to different federal states in Germany or even countries), our upfront training effort quickly amortizes, since a trained OutbreakFlow executes inference orders of magnitude faster than a case-based (non-amortized) method.
We now outline three training modes for OutbreakFlow.

\noindent
\\{\bf Simulation-Based Offline Learning: }
Traditional simulation-based approaches utilize a pre-computed reference table $\mathcal{D}^{(S)}$, which is a large data structure containing $S$ pairs $(\theta, X)$ of simulation parameters $\theta$ and corresponding synthetic observations $X$ \cite{abc1, abc2}. 
This strategy has also been used in machine learning approaches to simulation-based inference \cite{raynal2018abc,radev2020towards}, where the problem of inverse inference resembles a supervised learning task.
In the context of OutbreakFlow, the resulting \textit{offline learning} method is outlined in Algorithm \ref{alg:offline}.
It is particularly useful when calls to the simulator are computationally expensive: working with recorded synthetic data is then faster at the expense of higher memory demands during training. 

\begin{algorithm}[ht]
\caption{OutbreakFlow training phase using offline learning}\label{alg:offline}
\begin{algorithmic}[1]
\Require {$f$ - invertible inference network, $h$ - recurrent summary network, $g$ - convolutional filtering network, $\phi$ - neural network parameters, $S$ - total number of simulations, $B$ - number of simulations per batch (batch size).} 
\State {Generate a large reference table $\mathcal{D}^{(S)} := \{\theta^{(s)}, x_{1:T}^{(s)})\}_{s=1}^{S}$ by running the simulator $S$ times.}
\Repeat
\State{Sample a mini-batch of simulations: $\mathcal{D}^{(B)} \sim \mathcal{D}^{(S)}$.}
\State{Pass each simulated time-series through the filtering network: $\widetilde{x}^{(b)}_{1:T}=g(x^{(b)}_{1:T})$.}
\State{Pass the filtered time-series through the summary network: $y^{(b)}=h(\widetilde{x}_{1:T})$.}
\State{Pass each pair $(\theta^{(b)}, y^{(b)})$ through the inference network: $z^{(b)} = f(\theta^{(b)}, y^{(b)})$.}
\State{Compute loss from batch: $\mathcal{L}(\phi) =  - \sum_{b=1}^B\log q_{\phi}(\theta^{(b)} \given x^{(b)}_{1:T})$.}
\State{Update neural network parameters $\phi$ via backpropagation.}
\Until{convergence to $\phi^*$}
\State{\textbf{Return} trained networks $g, h, f$ with parameters $\phi^*$.}
\end{algorithmic}
\end{algorithm}

\noindent
\\{\bf Simulation-Based Online Learning:}
Instead of pre-computing synthetic data, we can generate a potentially limitless number of training pairs $(\theta, X)$ on-the-fly. 
Since each simulation result is discarded after the corresponding backpropagation update, the network never encounters the same inputs twice and overfitting is impossible. 
Moreover, the training phase can continue as long as the network keeps improving, as measured by continuous performance monitoring. 
Online training is outlined in Algorithm \ref{alg:online} and is used for all experiments in this work.
The present application lends itself to this approach, because the computational cost of running our epidemiological model is negligible, whereas more expensive simulations might become a bottleneck for this strategy.

\begin{algorithm}[ht]
\caption{OutbreakFlow training phase using online learning}\label{alg:online}
\begin{algorithmic}[1]
\Require {$f$ - invertible inference network, $h$ - recurrent summary network, $g$ - convolutional filtering network, $\phi$ - neural network parameters, $B$ - number of simulations per iteration (batch size).} 
\Repeat
\State{Generate a mini-batch $\mathcal{D}^{(B)} := \{\theta^{(b)}, x_{1:T}^{(b)})\}_{b=1}^{B}$ by running the simulator $B$ times.}
\State{Pass each simulated time-series through the filtering network: $\widetilde{x}^{(b)}_{1:T}=g(x^{(b)}_{1:T})$.}
\State{Pass the filtered time-series through the summary network: $y^{(b)}=h(\widetilde{x}_{1:T})$.}
\State{Pass each pair $(\theta^{(b)}, y^{(b)})$ through the inference network: $z^{(b)} = f(\theta^{(b)}, y^{(b)})$.}
\State{Compute loss from batch: $\mathcal{L}(\phi) =  - \sum_{b=1}^B\log q_{\phi}(\theta^{(b)} \given x^{(b)}_{1:T})$.}
\State{Update neural network parameters $\phi$ via backpropagation.}
\Until{convergence to $\phi^*$}
\State{\textbf{Return} trained networks $g, h, f$ with parameters $\phi^*$.}
\end{algorithmic}
\end{algorithm}

\noindent
\\{\bf Simulation-Based Hybrid Learning:}
Offline and online learning represent two extremes on a continuum of training strategies. 
Hybrid learning methods combine these two strategies and allow for a more fine-grained allocation of the available simulation budget. 
For instance, \cite{greenberg2019automatic} propose a round-based strategy, where each round incorporates its own simulation phase. 
Thus, the reference table is filled incrementally, and each round can reuse simulations from all previous rounds. 
Such a round-based training strategy is outlined in Algorithm \ref{alg:hybrid}}.

\begin{algorithm}[ht]
\caption{OutbreakFlow training phase using round-based hybrid learning}\label{alg:hybrid}
\begin{algorithmic}[1]
\Require {$f$ - invertible inference network, $h$ - recurrent summary network, $g$ - convolutional filtering network, $\phi$ - neural network parameters, $R$ - number of rounds, $S$ - number of simulations per round, $B$ - batch size.} 
\State{Initialize reference table $\mathcal{D}^{(R \times S)} := \{\}$.}
\For{$r = 1,...,R$}
\State{Generate synthetic data $\mathcal{D}^{(S)}_r :=\{\theta^{(s)}, x_{1:T}^{(s)})\}_{s=1}^{S}$ by running the simulator $S$ times.}
\State{Aggregate data: $\mathcal{D}^{(R \times S)} := \mathcal{D}^{(R \times S)} \cup \mathcal{D}^{(S)}_r$.}
\Repeat
\State{Sample a mini-batch of simulations: $\mathcal{D}^{(B)} \sim \mathcal{D}^{(R \times S)}$.}
\State{Pass each simulated time-series through the filtering network: $\widetilde{x}^{(b)}_{1:T}=g(x^{(b)}_{1:T})$.}
\State{Pass the filtered time-series through the summary network: $y^{(b)}=h(\widetilde{x}_{1:T})$.}
\State{Pass each pair $(\theta^{(b)}, y^{(b)})$ through the inference network: $z^{(b)} = f(\theta^{(b)}, y^{(b)})$.}
\State{Compute loss from batch: $\mathcal{L}(\phi_r) =  - \sum_{b=1}^B\log q_{\phi_r}(\theta^{(b)} \given x^{(b)}_{1:T})$.}
\State{Update neural network parameters $\phi$ via backpropagation.}
\Until{convergence to $\phi_r^*$}
\EndFor
\State{\textbf{Return} trained networks $g, h, f$ with parameters $\phi_R^*$.}
\end{algorithmic}
\end{algorithm}

\subsection*{\markchanges{Uncertainty Calibration and Computational Faithfulness}}

\markchanges{Computational faithfulness refers to the ability of a Bayesian method to recover the correct target posterior in a given modeling scenario.
It is an essential precondition for carrying out model-based predictions and interpreting the parameters of a model within a reference theoretical framework.
We can estimate the computational faithfulness of any BayesFlow application using simulation-based calibration \cite[SBC,]{talts2018validating}.
SBC is a diagnostic method which considers the performance of a sampling algorithm over the entire Bayesian joint model $p(\theta, X)$, regardless of the structure of the particular model.
It leverages the fact that most Bayesian models are generative by construction as well as the self-consistency of the Bayesian joint model. 
Accordingly, the \textit{average posterior distribution} over random draws from the Bayesian joint distribution $(\theta, X) \sim p(\theta, X)$ should always recover the prior distribution $p(\theta)$. \markchangesnew{In other words, for any given parameter combination $\theta^*$, the following should hold}:
\begin{equation}
    p(\theta^*) = \int \int p(\theta^* \given X)\,p(\theta, X)\,d\theta\,dX \label{eq:sbc}
\end{equation}
\markchangesnew{Random draws from $p(\theta, X)$ are generated by first sampling a configuration $\theta$ from the prior $p(\theta)$ and then running the (stochastic) simulator with the sampled parameter configuration to obtain a synthetic outbreak trajectory.
This process can be repeated multiple times and does not require an analytically tractable likelihood function.}
Importantly, if a Bayesian sampling method generates samples from the exact posterior, the equality implied by Eq.\ref{eq:sbc} should hold regardless of the particular form of the posterior. 
Thus, any violation of this equality indicates some error incurred by the sampling method.
The reasons for such an error can be either inaccurate
computation of the posterior or an erroneous implementation of the model itself \cite{talts2018validating}.

In practice, we approximate this integral by an ensemble of samples from many posterior distributions estimated from simulated time series with known generating parameters.
SBC uses a rank statistic (i.e., the number of posterior draws larger than the prior draw for each simulated time series) to compare the average posterior with the prior.
If Eq.\ref{eq:sbc} holds, then the rank statistic of each parameter will be uniformly distributed, allowing us to visually examine the equivalence using univariate histograms.
An inspection of the rank histograms thus provides a way to validate the computational faithfulness of OutbreakFlow within the scope of the modeling assumptions \cite{talts2018validating}.

A major disadvantage of SBC is that it can be extremely time-consuming, since it requires inverse inference on potentially thousands of simulated time series.
In addition, the obtained posterior draws should exhibit no autocorrelation for SBC to yield reliable results. The latter requirement makes it even more expensive when using MCMC or other non-amortized Bayesian methods yielding dependent samples.
Fortunately, amortized inference with OutbreakFlow alleviates these issues, since the inference phase on multiple time series is extremely efficient and posterior draws are independent given perfectly converged networks \cite{radev2020bayesflow}. 
Thus, validating the computational faithfulness of any BayesFlow application using SBC becomes a matter of seconds.}

\subsection*{Outbreak Prediction on the Basis of Estimated Posteriors}

\markchanges{The posteriors estimated by a trained OutbreakFlow network can be further used to make forecasts about the future dynamics of the pandemic, provided the parameters remain stationary. Due to changing intervention measures, population behavior, testing policies, and possible treatment advances, this is only true for a relatively short period beyond the observed time series, limiting the prediction horizon to a few weeks.
Given observed time series $X := x_{1:T}$, the posterior predictive distribution for upcoming data $X' := x_{T+1:T'}$ is given by:
\begin{equation}
p(X' \given X) = \int p(X' \given \theta, X)\,p(\theta \given X)\,d\theta    
\end{equation}
Although this quantity is hard to compute exactly, we can approximate it by running the simulation with parameters sampled from the posterior: $\{\theta^{(m)}\sim p(\theta\given X)\}_{m=1}^M$. 
Since the $\theta^{(m)}$ are drawn from the joint posterior, statistical dependencies and correlations between parameters are properly taken into account.
The resulting ensemble of $M$ simulated time series $\{\widetilde{X}^{(m)}\}_{m=1}^{M}$ can now be used to obtain point predictions (e.g., by computing their mean or median at each future time point) and to quantify the uncertainty of future scenarios (e.g., by computing quantiles or standard deviations at each time point).

In addition to future predictions, posterior predictive re-simulations are also crucial for further model validation.
If the estimated posteriors describe the real situation well, re-simulations should replicate the data that served to fit the model in the first place.
Mismatches between original and re-simulated time series indicate \textit{model misspecification} or a \textit{simulation gap}, that is, errors arising because important aspects of the disease dynamics or unknown corruption of the observed data are not properly represented by the model. 
These errors are invisible to simulation-based calibration, because it solely assesses whether the posteriors conform to model specifications.
Importantly, our OutbreakFlow experiments demonstrate very good agreement between original and re-simulated time series.}

\section*{Results}

\subsection*{OutbreakFlow as a \markchanges{Research Tool for Inferring} Dynamics of Emerging Epidemics}

In contrast to mainstream neural network applications, such as image or text analysis, analyzing the dynamics of emerging epidemics poses two major challenges: \markchanges{(i) data are usually sparse, with no large sets of training data available; and (ii) a reliable quantification of the uncertainty associated with neural network outputs, such as estimated parameters, is mandatory to allow for reliable subsequent evaluation of possible scenarios}.

Standard neural network architectures do not live up to these challenges. 
Therefore, we developed a novel method on the basis of a neural network architecture called BayesFlow \cite{radev2020bayesflow} that addresses the aforementioned challenges in two ways: (i) We leverage the epidemiological insight represented by SIR-type models by means of an alternative training procedure using simulated data -- \markchanges{{\it simulation-based training}; and (ii) we use networks that are specifically designed to perform Bayesian uncertainty quantification over their outputs.}

% This could be made a bit more technical (only 2-3 sentences)
In our framework, \markchanges{a large number of plausible hypothetical scenarios simulating the assumed epidemiological dynamics} is processed by the neural network until it becomes an expert in the interpretation of epidemiological observations. After completion of the training phase, the available real-world observations are passed to the network, which then estimates full Bayesian posterior distributions for the real-world parameters of interest. 
The design of our network architecture is depicted in Fig 1. The details of the method, as well as the model architecture are given in \textbf{Materials and Methods}.

The ultimate goal of our approach is comparable to that of traditional simulation-based Bayesian inference methods, such as ABC. However, our method operates much faster and generalizes, without retraining, to any real-world dataset within the scope of its training expertise \cite{radev2020bayesflow}.

\subsection*{\markchanges{Testing and Validation of OutbreakFlow}}

\markchanges{To validate our approach and test its performance in inferring parameter values in epidemiological models, we applied our architecture to a standard SIR-model describing the dynamics of an epidemic. 
This greatly simplified model is suitable for the initial two weeks of the pandemic and highlights essential properties of our approach.
It distinguishes between susceptible, $S$, infected, $I$, and recovered, $R$, individuals with infection and recovery occurring at a constant transmission rate $\lambda$ and recovery rate $\mu$, respectively. The model is defined by the following system of ODEs:
\begin{linenomath*}
\begin{align}
    \frac{dS}{dt} &= -\lambda\,\left(\frac{S\,I}{N}\right) \\
    \frac{dI}{dt} &= \lambda\,\left(\frac{S\,I}{N}\right) - \mu\,I \\
    \frac{dR}{dt} &= \mu\,I,
\end{align}
\end{linenomath*}
with $N = S + I + R$ denoting the total population number. 
In addition to the ODE parameters $\lambda$ and $\mu$, we consider a reporting delay \markchanges{parameter $L$} and a dispersion parameter $\psi$, which affect the number of reported infected individuals via
\begin{linenomath*}
\begin{equation}
    I_t^{(obs)} \sim \textrm{NegBinomial}(I^{(new)}_{t-L}, \psi),
\end{equation}
\end{linenomath*}
where $I_t^{(new)} = \lambda\,(S_t I_t/N)$} and we assume that the number of newly observed cases arises from a negative binomial distribution \cite{blumberg2014detecting}.
In addition, we estimate the initial number of infected individuals $I_0$, so the full parameter vector of interest becomes $\theta = (\lambda, \mu, L, \psi, I_0)$.
Priors over the five parameters are given in Table S4.

\begin{figure*}[ht]
\centering
\includegraphics[width=0.99\textwidth]{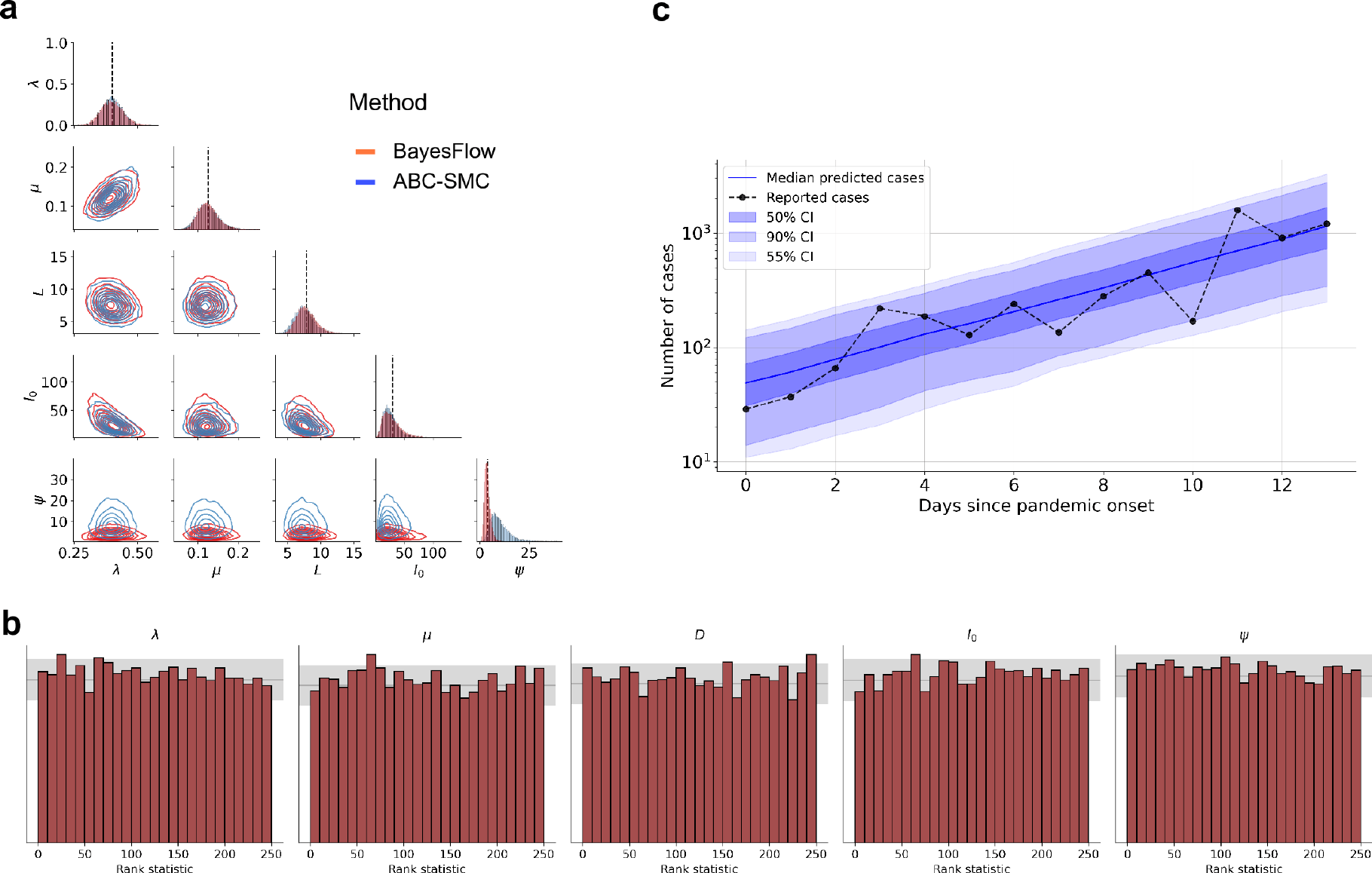}
\caption{\markchanges{(a) Bivariate posteriors over the five model parameters obtained by OutbreakFlow (red) and ABC-SMC (blue) from cases reported during the first 14 days of the Covid-19 pandemic in Germany. Results of the two methods are very similar except for the dispersion parameter $\psi$, where OutbreakFlow achieves superior uncertainty reduction (see text for a brief discussion); (b) Model-based predictions based on the posteriors obtained by OutbreakFlow; (c) Simulation-based calibration (SBC) computed from $10000$ simulated pandemic scenarios. The uniformity of the rank statistics indicates that there are no systematic biases in the approximate posteriors and implies trustworthy approximation.}}
\label{fig:validation}  
\end{figure*}

% With this validation study, we pursue the following goals. First, we want to validate the computational faithfulness of our BayesFlow architecture and demonstrate good posterior predictive performance on actually observed data. 
% For the latter, we utilize reported new cases from the first $14$ days of the Covid-19 pandemic in Germany.
% Second, we want to demonstrate that our results are consistent with those obtained via sequential Monte Carlo (SMC)-ABC \cite{klinger2018pyabc}. 
% Correspondingly, we apply the ABC-SMC algorithm on the observed time series to obtain $10,000$ posterior samples when either a maximum number of $12$ populations or a minimum tolerance $\epsilon < 0.1$ is reached. 
% We use the mean squared error between observed and simulated time series as a distance function. 
% Third, we want to show that the addition of non-identifiable parameters does not hurt the computational faithfulness or predictive performance of our architecture. 
% To test this, we augment the parameter vector with five dummy parameters $u_j \sim \textrm{Uniform}(0, 1)$ for $j \in \{1,2,3,4,5\}$ which are not involved in the data-generating (forward) model but are simply appended to $\theta$ during training.

\markchanges{As a first step, we trained our network on simulations from the simple SIR-type model formulated above above and then applied the network to the number of reported cases from the first 14 days of the Covid-19 pandemic in Germany. The results from this initial study are depicted in Fig 3. First, we observe that our posterior estimates are in line with those reported in a previous modeling study \cite{dehning2020inferring}, which utilized the same data and a similar model. 
Second, we note that the SBC plots indicate no systematic biases in the approximate posteriors and thus suggest that the posterior samples are trustworthy (assuming no simulation gap). 
Finally, the posterior predictive check indicates that the model can accurately reproduce the observed data (Fig 3).

\markchangesnew{Since Outbreak-Flow is especially designed to tackle complex stochastic models whose estimation usually necessitates simulation-based approaches, we compared its performance to an analysis performed by ABC-SMC, a popular approximate Bayesian computation algorithm based on sequential Monte Carlo sampling \cite{klinger2018pyabc}.
Our benchmark comparison reveals converging results.} However, OutbreakFlow implies a much sharper marginal posterior of $\psi$ than ABC-SMC, indicating more uncertainty reduction with regard to the dispersion parameter. 
Since the SBC plots indicate no overconfidence (i.e., no overdispersion) of the OutbreakFlow posteriors, it is likely that the ABC-SMC algorithm yields an underconfident (i.e., underdispersed) marginal posterior with respect to this one parameter. 
As for wall-clock running times, the ABC-SMC algorithm converged in $4.1$ hours, whereas OutbreakFlow trained with $50,000$ iterations using online learning required $4.3$ hours on the same laptop machine without parallel simulations.
This similar performance speaks in favor of amortized inference, as the training effort already amortizes after as a few as two data sets.

To further test if OutbreakFlow is able to provide a reliable quantification of model parameters, we analyzed if the additional consideration of non-identifiable parameters within the model would affect the calibration or predictive performance of our method. To this end, we included $5$ dummy variables $u_j \sim \textrm{Uniform}(0, 1)$ within our model that were not used for the generation of the simulated data, but were later included in the unknown parameter vector $\theta$ during training and inference. 

Performing the same training and inference phase with these $5$ additional dummy parameters neither hurts the calibration of OutbreakFlow, nor does it impact inference on the observed time series in a noticeable way, since the posterior estimates for the relevant parameters appear to be unaffected by the dummy parameters (see \textbf{Supplementary Information} for full results). The same is true for model-based posterior predictions, which underlines the ability of OutbreakFlow to reliably characterize parameter identifiability in case of insufficient data or over-parameterized models.} 

\subsection*{Inferring Epidemiological Characteristics from the Early Covid-19 Pandemic in Germany}

% Introduction for the analysis/ Model
\markchanges{After validation of the general applicability of our novel approach, we applied OutbreakFlow to data of the Covid-19 pandemic in Germany, analyzing reported cases (infected, recovered and deceased) in the time period from March 01, 2020 until June 11, 2020}. These data captured the early dynamics of the emerging epidemic associated with considerable uncertainty and stochasticity with regard to the number of detected cases, as well as the effect of subsequent public interventions. For our analysis, we used an extended SEIR-type model that had been developed recently and distinguishes between detected and undetected carriers of the disease comprising a total of 34 unknown model parameters (see Fig 2 and \textbf{Materials and Methods} for a detailed description of the model) \cite{khailaie2020estimate}. The model was trained on a time-period from March 01 until May 21 using wide prior distributions across plausible parameter ranges from previous literature \cite{dehning2020inferring, tang2020estimation}. The remaining data (three weeks from May 22 until June 11) are then used to assess the predictive value of the model.

% Results and Observations:
The observed and predicted dynamics, as well as the marginal posterior distributions of the individual model parameters are depicted in Fig 4 (\markchanges{see Fig S6 for simulation-based calibration}). Our model was able to recover the observed dynamics and yields good predictions for the future period, with its forecasts having well-calibrated uncertainty bounds for the newly infected, recovered, and diseased cases (see Fig 4). Despite the large number of unknown model parameters and limited data, our analysis indicated considerable reduction in uncertainty in relation to the prior knowledge for most of the model parameters (Fig 4). Standard point estimates (median, mean, MAP mode) and credibility intervals ($95\%$-CI between the $2.5\%$ and $97.5\%$ quantiles of the corresponding posterior) for all 34 model parameters are given in Table \ref{tab:tab4}.

% Parameter estimates
Interestingly, our parameter estimates (cf. Fig 4) are consistent with previous findings about central disease parameters \cite{dehning2020inferring}. With the number of undetected cases being one of the most important estimates to assess epidemic-related dynamics, our network estimates a median probability of remaining undetected (parameter $\alpha$) of $0.63$ with the maximum a posteriori (MAP) estimate at $0.79$ (95\%-CI [$0.07 - 0.91$]). Notwithstanding the large uncertainty surrounding the number of undetected cases, the posterior of $\alpha$ is clearly far from uniform (our prior assumption), and peaks well beyond $0.5$ (see Fig 4). 
\markchangesnew{This estimate is consistent with the results of representative seroprevalence studies in Germany \cite{harries2021sars,pritsch2021prevalence}, which find that $75\%$ of the sero-positive individuals at the end of the first wave had not been diagnosed with the disease before. An even higher fraction of undetected cases around $80\%$ was reported by seroprevalence studies focusing on the hotspot regions of Gangelt, Kupferzell, and Tischenreuth \cite{streeck2020infection,santos2020serology,wagner2021estimates}.}
Together with our estimated case fatality rate (parameter $\delta$) of $4.1\%$ (median) resp. $3\%$ (MAP), this results in an infection fatality rate of about $0.63\%$ (MAP estimates) or $1.5\%$ (median estimates). 

\begin{figure*}
\centering
\includegraphics[width=\textwidth]{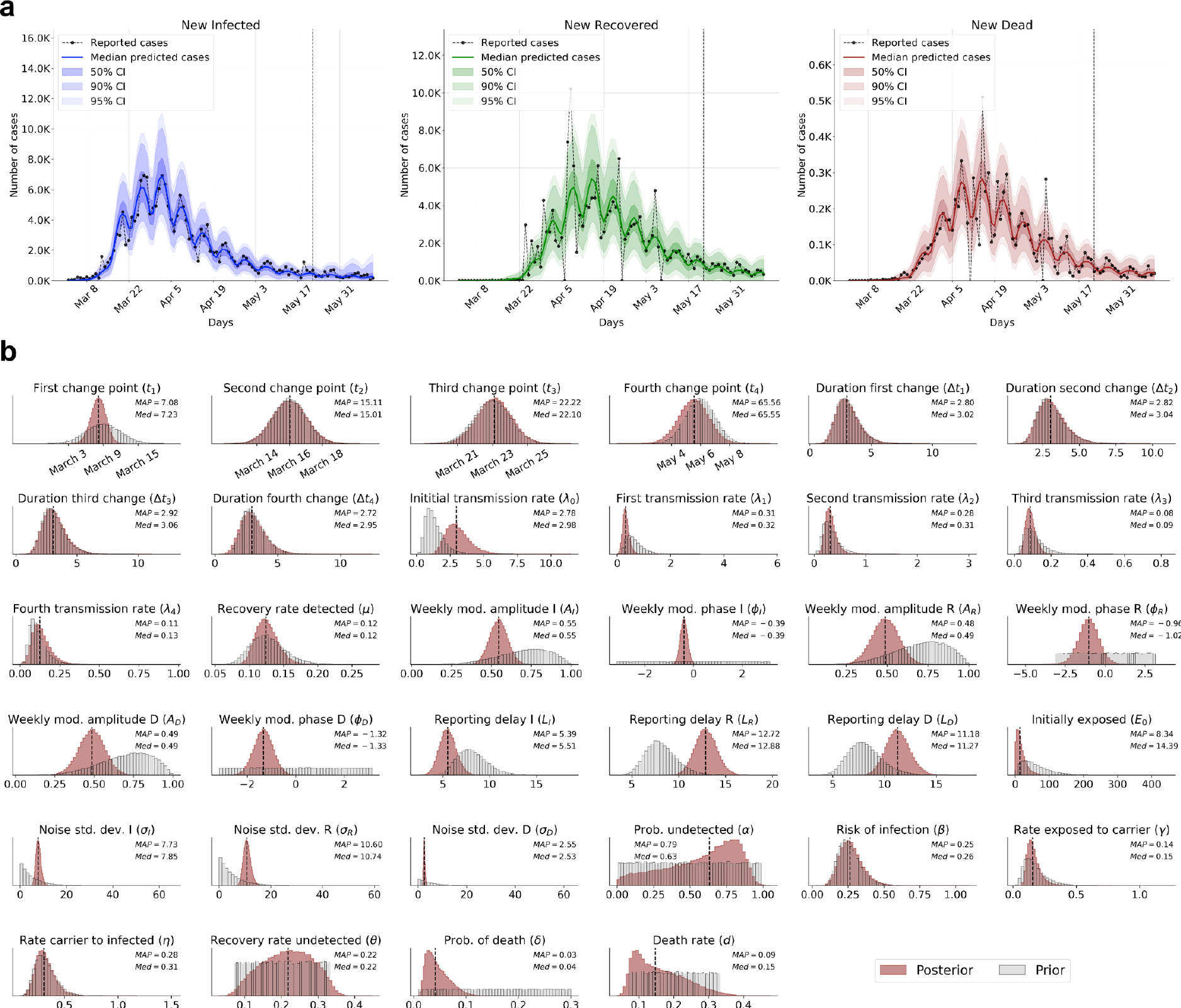}
\caption[short]{\markchanges{(a) Posterior predictions and forecasts of new cases obtained by inferring model parameters from epidemiological data available for reported infected, assumed recovered and deaths by Covid-19 for the entire Germany. Cases to the left of the vertical dashed line were used for posterior checking (model training) and cases to the right for posterior forecasts (predictions) on unseen data; (c) Marginal posteriors of all 34 model parameters inferred from data for the entire Germany alongside median and MAP summary statistics. Gray lines depict prior distributions for comparison with the posteriors. Vertical dashed lines indicate posterior medians.}}
\label{fig:entire}
\end{figure*}

% Incubation
\markchangesnew{Additionally, our informative priors for the parameters $\eta$ and $\gamma$ are not updated by the observed data.
Accordingly, around $3.23$ days will typically pass before the infection is detected (95\%-CI [$1.92 - 5.55$], $1/\eta$). This estimate is in line with results from \cite{guan2020clinical} (around $4$ days) and the World Health Organization \cite{world2020coronavirus} ($5-6$ days). 
Combined with the latent period (i.e., the time in compartment $E$: median $6.67$ days, 95\%-CI [$3.33 - 11.1$], $1/\gamma$), this leads to a median incubation period until symptom onset or a positive test outcome of around $9.9$ days, as originally implied by our priors.}

% Recovery
If the disease remains undiagnosed throughout (asymptomatic, weakly symptomatic etc.), recovery takes a median number of $4.59$ days (95\%-CI [$2.99 - 11.11$], $1/\theta$), a time period in which pre-symptomatic and undiagnosed individuals can be responsible for a considerable fraction of infections. Furthermore, if we assume that most infections occur near the end of the carrier stage, that is, after $2.5$ days in compartment $C$, we arrive at a generation time of around $9$ days. In conjunction with the delay for reporting infections $I$ of $5.5$ days (parameter $L_I$), this is consistent with the generally acknowledged fact that the success of intervention measures only becomes apparent after around two weeks. 

For diagnosed individuals, the median recovery period is estimated at $8.06$ days (95\%-CI [$6.13 - 10.20$], $1/\mu$).
Thus, manifestly ill cases have a more severe disease progression than undiagnosed individuals and typically require $1/\eta+1/\mu=3.23+8.06=11.3$ days until recovery \cite{gao2021systematic}. The time between diagnosis and death ($6.67$ days, 95\%-CI [$3.12 - 14.3$], parameter $1/d$) is shorter than in clinical reports, with parameter identification possibly impaired by the estimated reporting delay for disease-associated deaths $D$ of $11.3$ days (parameter $L_D$), which is probably much longer than in reality.
From the available time series alone, the model is not able to distinguish a long critical phase with short reporting delay from rapid death with long reporting delay. Nevertheless, it is remarkable how much information about $34$ free parameters our networks can extract from seeing only about $70$ time steps of real data (see Fig 4).

% Intervention methods
Finally, our results corroborate the timing of \markchanges{intervention measures} and the gradual reduction in transmission rate as observed in \cite{dehning2020inferring}. According to our estimates, the lifting of measures around May 6 would have led to an approximately $40\%$ increase in the transmission rate, as assumed by our prior. However, since the spreading rate at $t_4$ has already decreased to a median of $0.09$ (95\%-CI: [$0.05 - 0.15$]), the increase to a median of $0.13$ (95\%-CI [$0.05-0.28$]) does not lead to an exponential growth of infections.

\subsection*{\markchanges{Testing the Robustness of Model Analysis and Inferred Dynamics}}

\markchanges{To test the validity of our results, we performed a series of \textit{ablation studies} which independently reduce either the architecture of our neural network or the considered epidemiological model. 
Thereby, we are able to test (i) the importance of specific technical components for parameter inference, as well as (ii) the importance of specific aspects in the model structure when trying to explain the observed dynamics. 
Accordingly, the implemented changes comprise: : 1) removing the convolutional filtering network; 2) removing the recurrent summary network; 3) removing the observation model; 4) removing the intervention model; 5) removing the carrier ($C$) compartment from the latent disease model. 
For each of these ablation studies, we use the same number of simulations and training settings as before.
We then compared the predictive performance, calibration, and reliability in parameter estimation of each of the modified analyzes to our previous analysis.

The results from all ablation studies are available in \textbf{Supplementary Information}.
Indeed, our experiments indicate that all network and model components are crucial for the performance of OutbreakFlow and for the fidelity of model predictions.
Interestingly, no ablation scenario leads to dramatic miscalibration across all marginal posteriors.
However, predictive performance markedly deteriorates in all ablation studies.
For instance, removing either the intervention or the observation model leads to the rejection of all time series drawn from the posterior predictive due to divergences. 
Further, removing the latent carrier compartment leads to poor fits and notable underestimation of the reported cases. 
On the other hand, removing either the filtering or the summary network does not lead to dramatic misfits but
prevents the architecture from fully capturing the structure of the observed time series (e.g., weekly modulation is poorly captured, see \textbf{Supplementary Information}).}

\begin{table}
\caption{Posterior summaries and 95\%-CIs for each model parameter inferred from data for the entire Germany.}
\centering
\begin{tabular}{llllll}
\toprule
Parameter & Symbol &  Median &    Mean &     MAP &              95\%-CI \\
\midrule
Onsets of changes & $t_1$  &  March 8  &   March 8 &   March 8 &    [March 6 - 11] \\
 &  &  (Day 7.23)  &   (Day 7.20) &   (Day 7.08) &    [Day 4.64 - 9.59] \\
   & $t_2$  &  March 16  &   March 16 &   March 16 &    [March 14 - 18] \\
 &   &  (Day 15.01) &  (Day 15.01) &  (Day 15.11) &  [Day 12.99 - 17.05] \\
   & $t_3$  &  March 23  &   March 23 &   March 23 &    [March 21 - 25] \\
 &   &  (Day 22.10) &  (Day 22.10) &  (Day 22.22) &  [Day 20.23 - 24.00] \\
   & $t_4$ &  May 6  &   May 6 &   May 6 &    [May 4 - May 8] \\
 &     &  (Day 65.55) &  (Day 65.53) &  (Day 65.56) &  [Day 63.48 - 67.54] \\
Duration of changes $\,\,[\text{days}]$ &  $\Delta t_1$ &   3.02 &   3.14 &   2.80 &    [1.61 - 5.34] \\
 & $\Delta t_2$ &   3.04 &   3.16 &   2.82 &    [1.65 - 5.34] \\
 & $\Delta t_3$ &   3.06 &   3.18 &   2.92 &    [1.63 - 5.41] \\
 & $\Delta t_4$ &   2.95 &   3.08 &   2.72 &    [1.45 - 5.43] \\
Transmission rates & $\lambda_0$  &   2.98 &   3.12 &   2.78 &    [1.70 - 5.31] \\
 & $\lambda_1$  &   0.32 &   0.34 &   0.31 &    [0.13 - 0.63] \\
 & $\lambda_2$  &   0.31 &   0.33 &   0.28 &    [0.16 - 0.58] \\
 & $\lambda_3$  &   0.09 &   0.09 &   0.08 &    [0.05 - 0.15] \\
 & $\lambda_4$  &   0.13 &   0.14 &   0.11 &    [0.05 - 0.28] \\
\midrule
Reporting delays & $L_I$        &   5.51 &   5.54 &   5.39 &    [3.87 - 7.35] \\
 & $L_R$        &  12.88 &  12.91 &  12.72 &  [10.73 - 15.21] \\
 & $L_D$        &  11.27 &  11.31 &  11.18 &   [9.15 - 13.65] \\
Weekly modulation amplitudes & $A_I$        &   0.55 &   0.55 &   0.55 &    [0.43 - 0.66] \\
 & $A_R$        &   0.49 &   0.49 &   0.48 &    [0.33 - 0.65] \\
 & $A_D$        &   0.49 &   0.49 &   0.49 &    [0.32 - 0.64] \\
Weekly modulation phases & $\phi_I$     &  -0.39 &  -0.39 &  -0.39 &  [-0.69 - -0.09] \\
 & $\phi_R$     &  -1.02 &  -1.02 &  -0.96 &   [-2.36 - 0.33] \\
 & $\phi_D$     &  -1.33 &  -1.33 &  -1.32 &  [-2.13 - -0.55] \\
Reporting noise scales & $\sigma_I$   &   7.85 &   7.92 &   7.73 &   [5.96 - 10.31] \\
 & $\sigma_R$   &  10.74 &  10.85 &  10.60 &   [8.36 - 13.88] \\
 & $\sigma_D$   &   2.55 &   2.54 &   2.53 &    [1.95 - 3.23] \\
\midrule
Number of initially exposed & $E_0$        &  14.39 &  18.72 &   8.34 &   [1.20 - 61.37] \\
Risk of infection from $I$ & $\beta$       &   0.26 &   0.27 &   0.25 &    [0.14 - 0.45] \\
Rate $E \rightarrow C\qquad[1 /  \text{days}]$& $\gamma$     &   0.15 &   0.17 &   0.14 &    [0.09 - 0.30] \\
Rate $C \rightarrow I\qquad\,\,[1 /  \text{days}]$ & $\eta$       &   0.31 &   0.32 &   0.28 &    [0.18 - 0.52] \\
Rate $I \rightarrow R\qquad\,\,[1 /  \text{days}]$ & $\mu$        &   0.12 &   0.13 &   0.12 &    [0.10 - 0.16] \\
Rate $C \rightarrow R\qquad[1 /  \text{days}]$ & $\theta$     &   0.22 &   0.22 &   0.22 &    [0.09 - 0.33] \\
Rate $I \rightarrow D\qquad\,[1 /  \text{days}]$ & $d$          &   0.15 &   0.16 &   0.09 &    [0.07 - 0.32] \\
Probability of $C \rightarrow R$ & $\alpha$      &   0.63 &   0.58 &   0.79 &    [0.07 - 0.91] \\
Probability of $I \rightarrow D$ & $\delta$     &   0.04 &   0.05 &   0.03 &    [0.02 - 0.10] \\
\bottomrule
\end{tabular}
\label{tab:tab4}
\end{table}

\subsection*{Predicting the Epidemiological Dynamics within the Individual German Federal States}

In the previous section, we demonstrated the ability of our method to recover observed epidemiological dynamics and to infer reliable parameter estimates that determine disease characteristics based on data covering the early Covid-19 epidemic in whole Germany. 
Arguably, the importance of uncertainty concerning the number of reported cases, as well as stochastic effects increase with lower case numbers. 
To this end, we also applied our method to each German federal state separately, as individual states are characterized by different population sizes, as well as different onsets and progression of the epidemic. 
\markchanges{Because data for recovered cases per federal state was lacking (in contrast to recovered cases for the entire Germany), the network was trained solely on the simulated reported infected cases and deaths.}
Note that we only trained a \textit{single} OutbreakFlow, which we then applied unchanged to each German federal state.
In this way, the training effort amortized over the repeated applications of the same neural estimator (see Fig S13 for simulation-based calibration of the trained network).

% Predictions and uncertainty
Posterior predictions and forecasts for cumulative infections (\markchanges{derived from estimated new cases}, see Fig S14) in each federal state are depicted in Fig 5 (see also Fig S15 for predictions of cumulative deaths). As for Germany as a whole, we observe that median predictions follow very closely the reported cumulative number of cases across all federal states. Furthermore, the reported cases are very well represented by the uncertainty bounds derived from the parameter posteriors, with prediction uncertainty increasing towards the future (i.e., predictions after the dotted vertical lines in Fig 5). However, median predictions can become unreliable when only a few cases are available for model training (see predictions for cumulative deaths for the state Mecklenburg-Western Pomerania in Fig S15). Therefore, well-calibrated uncertainty estimates are particularly important and need to be taken into account when reporting point predictions.

% Individual parameters/ Discussion
The posterior distributions of individual model parameters for each of the 16 German federal states are depicted in \textbf{Supplementary Information}. 
For comparison of individual estimates between different states, we focused on four latent parameters that are essential for assessing early epidemical dynamics: (i) the probability of infecteds to remain undetected ($\alpha$), (ii) the recovery rate of undetected ($\theta$), (iii) the number of initially exposed individuals ($E_0$), and (iv) the initial transmission rate ($\lambda_0$).

\begin{figure*}
\centering
\begin{subfigure}{.99\textwidth}
    \includegraphics[width=\textwidth]{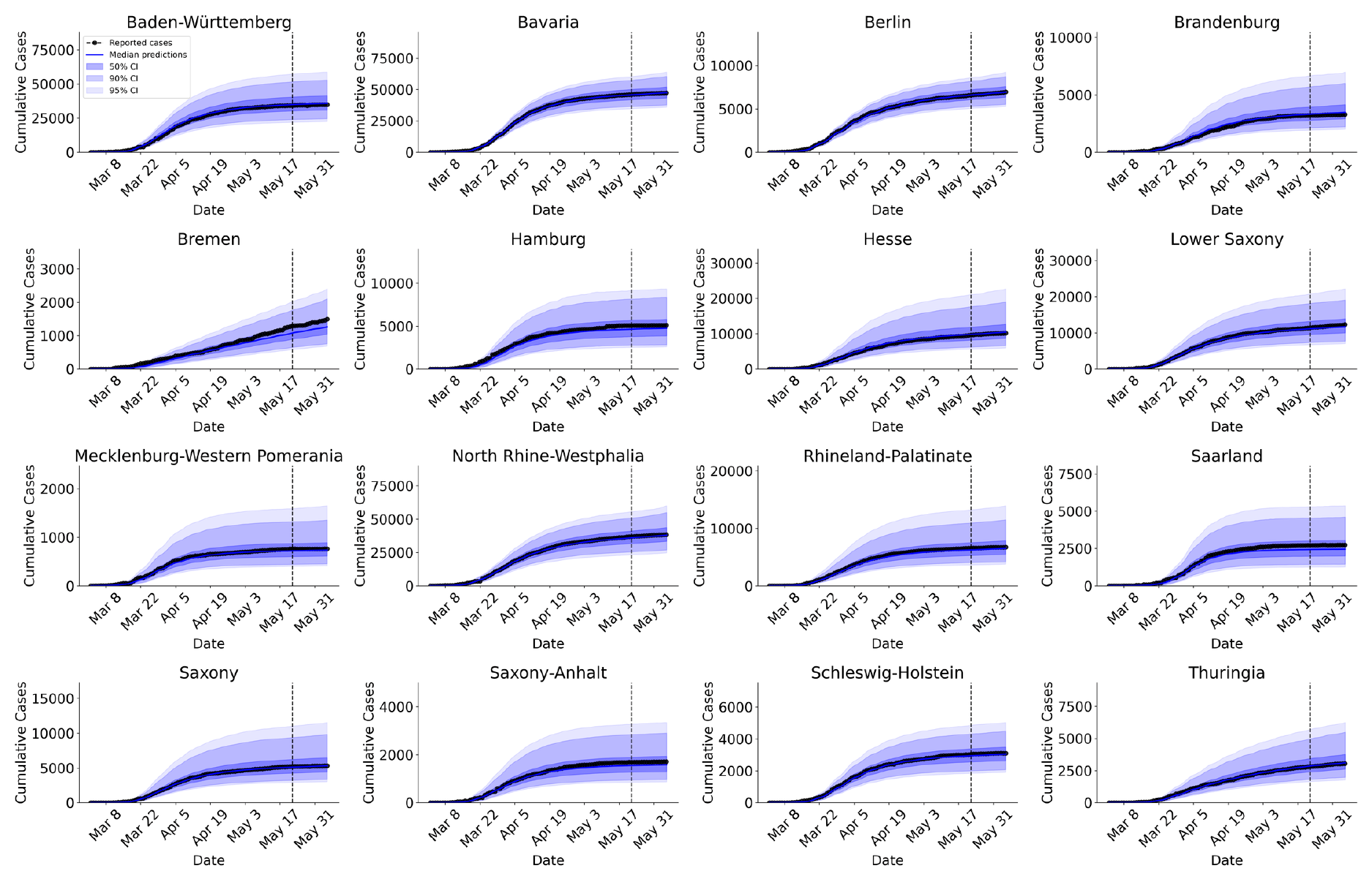}
\end{subfigure}
\caption[short]{Model predictions of cumulative Covid-19 cases (\markchanges{derived from estimated new cases}) for each German federal state. Cases to the left of the vertical dashed line (8 weeks) were used for model fitting and posterior checking and cases to the right (3 weeks) for forecasts on new data. We observe that median model predictions closely match both past and future reported cases for each German federal state. Most importantly, the reported cases always lie within the estimated CIs, which vary across the federal states.}
\label{fig:states}
\end{figure*}

\begin{figure*}
\centering
\includegraphics[width=\textwidth]{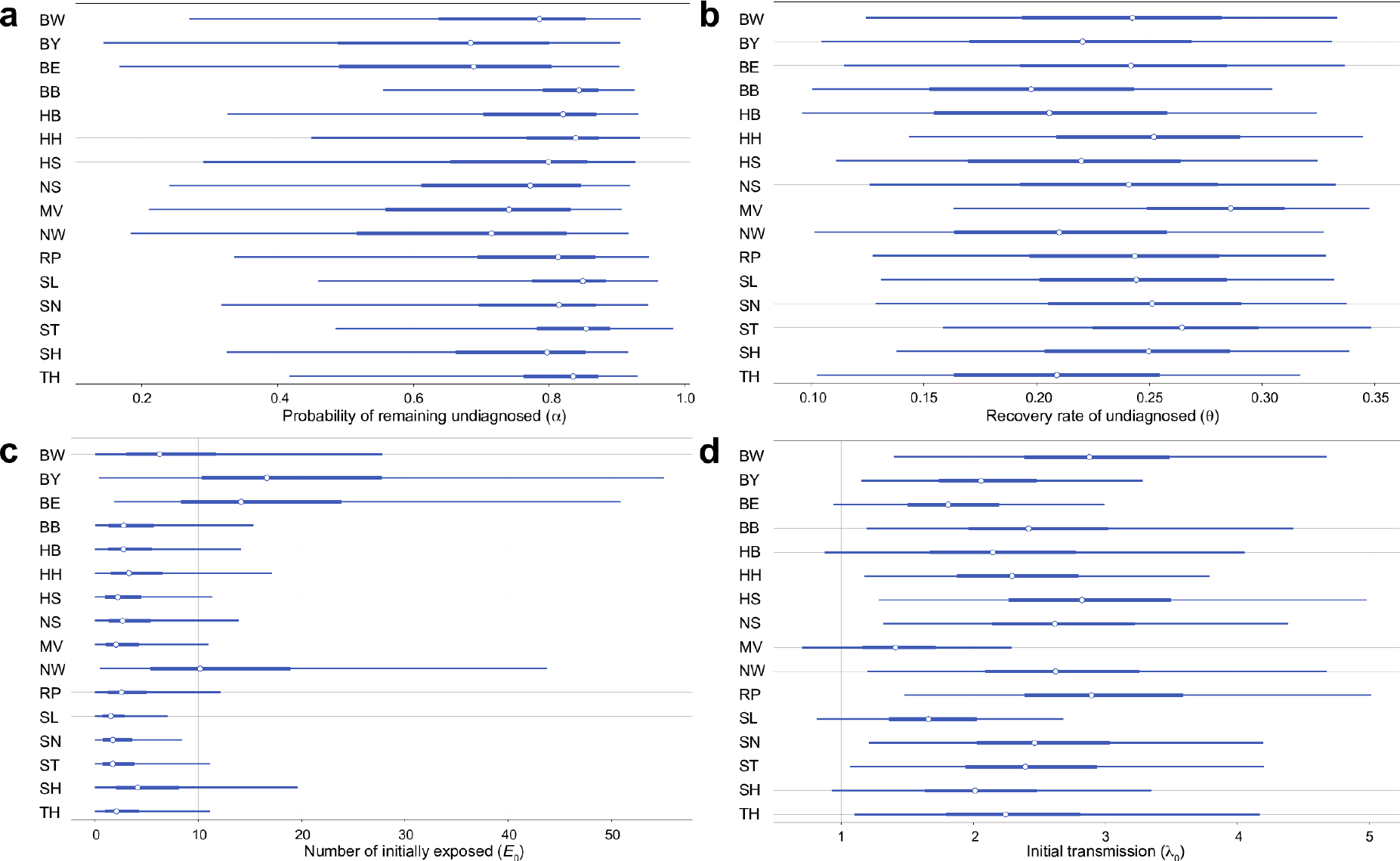}
\caption[short]{\textbf{(a)} Forest plot depicting 95\% credibility intervals for the probability of remaining undetected ($\alpha$) obtained by amortized inference on data from all German federal states. Thin lines depict highest posterior density (HDI) intervals, thick lines depict posterior quartiles, and white points depict the corresponding medians of the estimated posteriors; \textbf{(b-d)} Corresponding plots for \textbf{(b)} the recovery rate of undetected ($\theta$), \textbf{(c)} the number of initially exposed ($E_0$), and \textbf{(d)} the initial transmission rate ($\lambda_0$).}
\label{fig:forests}
\end{figure*}

We observe that posterior estimates of $\alpha$ across states tend to peak well above $0.5$ (see Fig 6a), \markchanges{suggesting once again that many infections have remained undetected/undiagnosed during the initial months of the Covid-19 pandemic in Germany}. Furthermore, we observe that some states have smaller probabilities of undiagnosed infections, especially Bavaria, Berlin, and North Rhine-Westphalia (\markchanges{possibly indicating a more successful testing policy}), although these estimates are associated with \markchanges{substantial} uncertainty. Interestingly, the $\alpha$ posteriors of each  state are considerably sharper than the posterior of $\alpha$ for the entire Germany (see \textbf{Supplementary Information} for full marginal posterior plots for each state). \markchanges{This may be a consequence of the high variability of $\alpha$ across states}, but could also be an estimation artifact of the simplified model (only two epidemiological time series, new infections and deaths, are available for the states).  

Less \markchanges{variability between states} is observed in the estimates for the recovery rate of undetected, $\theta$, suggesting an overall fast recovery of undiagnosed individuals (Fig 6b). However, some \markchanges{differences between states} are evident in the estimates of initially exposed individuals, $E_0$ (cf. Fig 6c), \markchanges{with the states of Bavaria, Berlin, and North Rhine-Westphalia having significantly more exposed individuals at onset than other states, which may correspond to observed outbreaks in these states after skiing holidays and festivals}. Finally, Fig 6c depicts a comparison of initial transmission rates $\lambda_0$ between states. We observe that estimates vary around a median value of $2.27$ across states, with the state Mecklenburg-Western Pomerania having the lowest and the state Baden-W\"{u}rttemberg having the highest median transmission rate at onset.

% Summary
Thus, our analysis is able to reveal differences in the epidemiological dynamics between individual German federal states and proves reliable for different population sizes.

\section*{Discussion}

% Brief summary
In this work, we presented a novel simulation-based Bayesian inference framework for complex epidemiological models. 
We directly demonstrated the utility of our method by applying it to publicly available data on the early reported infected, recovered, and deceased individuals during the first phase of the Covid-19 pandemic in Germany.

% Usefulness of method given limited data and uncertainty quantification
Given the general uncertainty in reported numbers for emerging infectious diseases, estimation methods need to account for this uncertainty when providing parameter estimates and be able to efficiently incorporate incoming data. 
Our method works reliably well, providing well calibrated uncertainty bounds for individual parameter estimates even in case of small sample sizes and a limited amount of observations. Given the limited amount of data available for model calibration (i.e., number of reported infections, defined recoveries and deaths) and the uncertainty concerning these values, our analysis is able to \markchangesnew{reduce the uncertainty about} quantities that could not be obtained during the time course of the epidemics.

% Parameter estimates of undetected
\markchangesnew{Our estimates suggest that a large fraction of the infected individuals (60-80\%) might have gone undetected through the course of the early Covid-19 outbreak in Germany. This finding has been confirmed by subsequent seroprevalence studies from Germany \cite{harries2021sars,pritsch2021prevalence,streeck2020infection,santos2020serology,wagner2021estimates} and other countries \cite{xu2020seroprevalence,pollan2020prevalence,stringhini2020seroprevalence,knabl2021high}
and is also in line with estimates on reporting rates \cite{pullano2021underdetection}}.
However, our posteriors also suggest that there is non-negligible uncertainty surrounding this estimate when derived in a purely model-based manner. Moreover, different summary statistics (e.g., means, medians, MAPs) derived from non-symmetric posteriors offer slightly different conclusions. This observation highlights the need to consider the full posteriors and corresponding credibility intervals when aiming to draw substantive conclusions about epidemiological parameters and possible forecasts for the progression of the epidemic or the effect of specific public health interventions.  
% Parameter interpretation
When interpreting the results of parameter estimates, one should also be aware that mechanistic models like the one used here only describe the average behavior of entire compartments. Accordingly, the given confidence intervals quantify our uncertainty about the inferred parameter averages and can not serve as a measure for the variability between individual cases.

% General applicability & Outlook
Our approach has two key advantages over standard Bayesian and \markchanges{likelihood-based} point estimation methods. 
First, it can flexibly deal with arbitrarily complex models and data structures, requiring no closed-form likelihoods or \textit{ad hoc} distributional restrictions regarding the shape of the joint prior or posterior. As standard SIR-models based on (stochastic) ordinary differential equations generally provide a coarse-grained view on the epidemic dynamics \cite{keeling2011modeling}, more complex models accounting for heterogeneous social interactions, age-dependent effects, and/or spatial and temporal heterogeneity become more and more important to predict the progression of an epidemic or guide \markchanges{intervention measures} \cite{neher2020potential, baker2020susceptible, ferguson2006strategies,davies2020effects, dehning2020inferring, flaxman2020estimating}. Such agent-based and stochastic models can be easily incorporated within our neural Bayesian framework.

As a second advantage, \markchanges{the \textit{amortized inference} property of our method, that is, training the network only \textit{once} on simulated data, allows efficient posterior sampling and simultaneous application to multiple data sets}. In addition, it allows for efficient online-learning and validation (i.e., the continuous integration of upcoming data), once the networks have been trained with sufficient amounts of simulated data. These advantages are important, since they enable researchers to concentrate on formulating, testing, and validating complex model systems without worrying about estimation efficiency or analytical tractability. 

% Future Developments and Caveats
Future developments will include Bayesian model comparison, multilevel modeling with hierarchical priors and a systematic comparison between different neural inference architectures. 
Hierarchical modeling allows us to better distinguish between disease-specific and region-specific parameters by representing the natural cluster structure of epidemiological data. 
In addition, model comparison is an especially important research avenue to compare different possible disease transmission and progression schemes \markchanges{that could explain the observed dynamics}.
Currently, in order to compare multiple competing dynamic models, separate neural networks would need to be trained and stored - one network corresponding to each model.
Even though such a training can be carried out in parallel, it would be much more efficient to embed all competing models within a single neural architecture, which can perform both prior predictive and posterior predictive model comparison. 
This idea has already been explored in \cite{radev2020amortized} and future research should investigate its utility for comparing complex epidemiological models.

\markchangesnew{In addition, tackling non-stationary dynamic processes is an important open problem of our framework that we plan to address in future work. This limitation is one reason why we focus on the early outbreak dynamics, where the assumption of stationarity appears plausible (e.g., absence of major virus mutations, largely fixed testing policy and implied fraction of undetected infections, no significant progress on treatment options and subsequent recovery times), with the exception of government interventions and behavioral changes, which we explicitly model. 
Moreover, our Bayesian treatment ensures that deviations from stationarity and shortcomings of the reported data do not result in catastrophic failure of our method, but are reflected in wider uncertainty regions than would otherwise have been achievable.}

\markchanges{When applying OutbreakFlow, one should be aware of three potential error sources that can distort the outcomes and interpretations of amortized Bayesian workflows.
First, model misspecification and data contamination can result in a \textit{simulation gap}.
A simulation gap occurs when the model cannot represent the actual disease dynamics or when data collection is biased or contaminated in ways not accounted for by the model.
We addressed these issues by suitable model extensions, which are motivated by theoretical considerations and ablation studies.
Remaining misspecifications can often be detected via standard Bayesian model checking methods, for instance, insufficient posterior predictive accuracy, divergent re-simulations, or very low posterior probability under the prior.
However, theoretical guarantees on upper bounds for the residual errors remain an important open problem of our approach.

The second source is the \textit{Monte Carlo error} introduced by approximating the expectation in Eq.\ref{eq:nll} with only a finite number of simulations.
It is also referred to as approximation error and widely acknowledged throughout all Monte Carlo methods. 
This error can be mitigated relatively easily in our online learning setting, because we can generate a potentially endless stream of synthetic training data until the continuously monitored prediction accuracy is satisfactory. 
In this respect, simulation-based inference is better positioned to fully utilize the capacity of deep neural networks than traditional supervised learning methods, which rely on a limited supply of annotated real data.

The third source of error is an \textit{amortization gap}, which refers to potential deficiencies of the network for atypical real dynamics.
Atypical dynamics are, by definition, underrepresented in any training set, and an amortized inference scheme may be less accurate in such cases.
An amortization gap can be detected via the probabilistic calibration methods advocated in this work, that is, simulation-based calibration. 
When the amortization gap is too large, resorting to non-amortized methods such as \cite{papamakarios2019sequential,greenberg2019automatic} might be a viable option. 
Alternatively, the accuracy of an OutbreakFlow for more rare situations can by improved by increasing the expressiveness of the filtering, summary, and inference networks (e.g., more layers and more units,  along with more training iterations). 
The choice of optimal neural network architectures and training algorithms is a topic of ongoing research in the deep learning community and an important target for future work on simulation-based inference.}

% Final summary
In summary, our \markchanges{OutbreakFlow} architecture provides a general inference framework for complex epidemiological scenarios and enables an efficient simulation-based inference for key epidemiological parameters. 
We therefore believe that our proposed architecture can facilitate uncertainty-aware inference with complex realistic epidemiological models, \markchanges{especially during the early phase of epidemic outbreaks when information is scarce and data reliability low. 
Rapid trustworthy inference can timely reveal crucial dynamic aspects of a spreading disease and inform effective public health interventions}.

\bibliography{references}
\bibliographystyle{plos2015.bst}

\end{document}